\documentclass[epsf]{ptptex}
\usepackage{graphicx}
\inst{Department of Earth and Planetary Sciences, Kobe University, Kobe 657-8501, Japan}
\recdate{April 23, 2004}
\abst{Clarke and Pringle (2004) derived a proper viscosity formula in a rotating 
gas by applying mean free path theory. We study their argument in detail and 
show that their result can be derived with a much simpler calculational 
procedure and a physically clearer picture.  
}
\begin{document}

\title{On kinetic theory viscosity in a rotating gas }
\author{Takuya {\sc Matsuda} and Eiji {\sc Hayashi} }
\maketitle

\section{Introduction}

The subject of angular momentum transport is important in theories 
concerning
accretion disks. The $\alpha$ model of Shakura and Sunyaev\cite{rf:S-S}
assumes outward transportation of angular momentum due to some viscous
force. This viscosity is believed to originate from turbulence, because
the molecular viscosity is far too small in this case.

Because the viscosity formula for turbulence is not well known, generally accepted
practices have relied on the formula for molecular viscosity, with the
viscosity coefficient assumed to be far larger than that for actual molecular viscosity.\cite%
{rf:S-S} In the present paper we investigate the molecular viscosity formula,
which can be unanbiguously defined, applicable to a rotating gas. The viscosity
formula for shear flow is well known. The shear stress is proportional to
the rate of strain. For a shear flow represented by $\mbox
{\boldmath $u$} = (0, U(x))$%
, the $x$-$y$ component of the shear stress is given by \newline
\begin{equation}
\sigma_{xy}=-\eta \left( dU/dx\right),   \label{linear}
\end{equation}
where $\eta$ is the viscosity coefficient.

We next consider a rotating flow represented by 
$\mbox{\boldmath $u$} =(0, R\Omega(R))$
in cylindrical coordinates, where $R$ and $\Omega$ are the radial distance and
angular velocity, respectively. In this case the $R$-$\phi$ component of the
shear stress is \newline
\begin{equation}
\sigma_{R\phi}=-\eta R(d\Omega/dR).   \label{circular}
\end{equation}

Microscopically, the formula (\ref{linear}) can be derived with
mathematical rigor from the Boltzmann equation using the Chapman-Enskog
expansion.\cite{rf:V-K} It can also be derived more readily by applying the
mean free path theory of the kinetic theory of gases
heuristically.\cite{rf:V-K,rf:RL}
 A simple application of the mean free path theory to a rotating
gas, however, leads to the incorrect result $\sigma_{R\phi} \propto -
d(R^2 \Omega)/dR$.\cite{rf:H-M} The correct derivation starting from the
Boltzmann equation naturally leads to the formula (\ref{circular}).
\cite{rf:K-I}\tocite{rf:H} It is a puzzle why the 
application of
the mean free path theory, which seems physically plausible, does 
not yield
the correct result.

Recently, Clarke and Pringle\cite{rf:C-P} have shown that application of
the mean free path theory to a rotating gas can indeed lead to the correct
formula, (\ref{circular}). They consider the case in which
the shear velocity of the gas is sufficiently smaller than the thermal
velocity of the molecules. Their argument employs the inertial frame and
approximates molecular orbits by straight lines.

We have been inspired by Ref. 8) and, focussing on
that paper, show here that their result can be derived with a simpler
calculation procedure and a physically clearer picture. 

\section{Non-rotating, linear shear flow}

We consider, for simplicity, an inertial system in two dimensions, 
in which there exists a
parallel flow in the $y$ direction.  We consider a
point S in the flow and define
its coordinates as $(x_0, 0)$ (see Fig. 1 with $U^{\prime}<0$, 
where $U^{\prime}$ is the velocity gradient of the flow). An
observer moves with S, and our argument employs his rest frame, which is
also an inertial frame. As seen by the observer, the velocity of the flow can
be written $\Delta{\mbox{\boldmath $u$}}
=(0,(x-x_0)U^{\prime})$. Let us consider a point E that 
is at a distance $\lambda/2$
from S, where $\lambda$ is the mean free path of gas molecules.\footnote{%
Although Clarke and Pringle chose this distance as $\lambda$, we consider
S to be the midpoint of the trajectory of a molecule and take this
distance to be $\lambda/2$.}
Hereafter, we
assume that the flight lengths between 
succesive collisions for all molecules are
the same, i.e. $\lambda$, for simplicity. 

The velocity of the flow at the point E as seen from S is directed in the
$y$ direction, and its magnitude can be written as

\begin{equation}
\Delta u=-\frac{1}{2}\lambda U^{\prime}\cos \alpha,   \label{deltau}
\end{equation}
where $\alpha$ denotes the angle between the lines SE and SA.
The flow velocity at A is also $\Delta u$, as is clear from
Fig. 1.  The situation here is similar to that in Ref. 8).

\begin{figure}[tbp]
\begin{center}
\includegraphics[height=11cm]{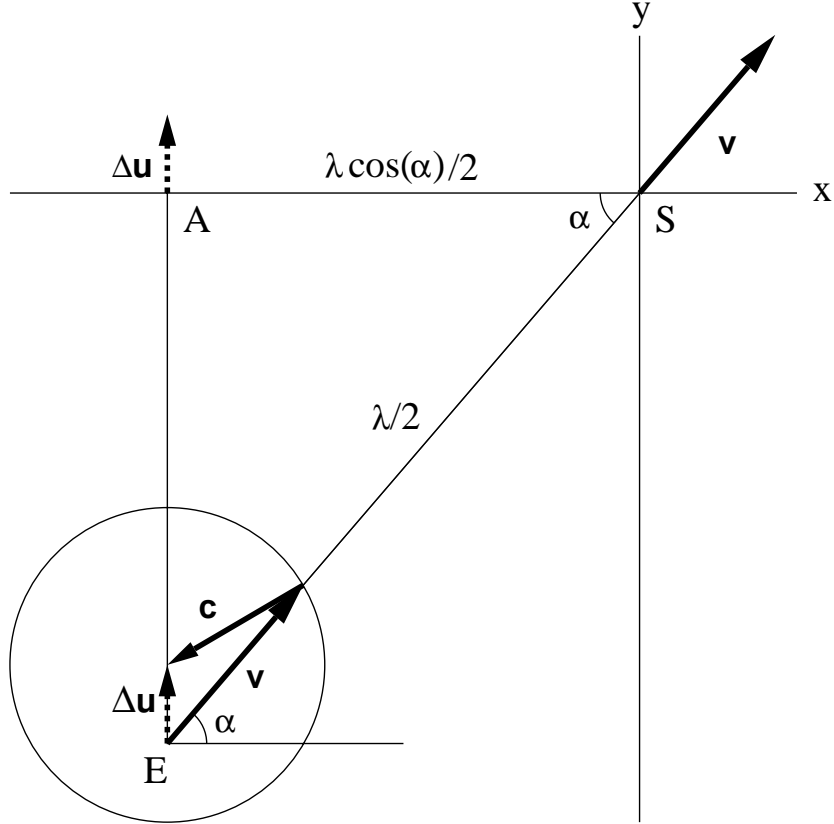}
%\centerline{\epsfbox{draw.eps}}
\end{center}
\caption{Schematic diagram showing velocities of gas flow and a gas
molecule. Here, E is the emission point, S the observational point, 
c the thermal velocity of
a molecule, $\Delta u$ the mean flow velocity at E and A, $\protect\lambda$
the mean free path, and $\protect\alpha$ the 
angle between SE and SA. In this figure $%
U^{\prime}<0$. }
\label{fig1}
\end{figure}

Gas molecules are assumed to be ejected from the point E isotropically when
observed from the frame moving with E, and at constant speed $c$.\cite{rf:C-P}
The flow velocity and molecular velocity added give $v$.
Application of cosine theorem to the triangle formed by $\Delta u, c \,\, %
\mbox{and}\, v$ gives 
\[
c^2 = v^2 +(\Delta u)^2 - 2v \Delta u \cos\left (\frac\pi2-\alpha
\right ). 
\]
Following Ref. 8), we assume $c^2 \gg (\Delta u)^2$,
ignore higher-order terms, and obtain 
\begin{equation}
v=c\left (1-\frac{1}{2}\frac{\lambda U^{\prime}}{c}\cos \alpha \sin \alpha
\right ) =c \left (1-%
\frac{1}{4}\frac{\lambda U^{\prime}}{c}\sin2\alpha \right ).   \label{v}
\end{equation}
Gas molecules ejected from the point E pass through the point S without
a change in their velocities or directions of motion. Equation (4) 
shows that the
velocity distribution of the gas molecules at S is no longer isotropic but
exhibits an oval shape whose major axis is inclined \(45^{\circ}\)
toward the $x$-axis.

The $x$-$y$ component of the viscous stress tensor $\sigma_{xy}$ is the net
$y$ momentum, carried by the gas molecules through a line of unit length
along the $y$-axis per unit time. The $x$ and $y$ components of the velocity are 
$v_{x}=v\cos \alpha$ and $v_{y}=v\sin \alpha$, respectively. The $y$ momentum
is $m v_y$, and the mass flux is $n v_x$, where $m$ and $n$ are mass of a
molecule and the number density of molecules, respectively. Thus, $%
\sigma_{xy}=mn\langle v_{x}v_{y}\rangle$, where $\langle 
\cdot\cdot\cdot\rangle$ represents an
average taken over velocity space.

Because we assume that the molecule velocity at E has a constant value $c$
in all directions,
the distribution function is non-zero only on the circle with radius $c$, and
therefore it must have the form $f(v) \sim \delta(v-c)$, where $\delta$ is
the delta function and
$v$ is the radial coordinate in velocity space. Integration of $\delta(v-c)$
over $v$-$\alpha$ space yields $2 \pi c$.
Then, using $v$ in Eq. (\ref{v}), our distribution function is 
found to be
\begin{equation}
f(v, \alpha)=\frac{1}{2\pi c}\delta 
\left ( v-c \left ( 1-\frac{1}{4}\frac{\lambda U^{\prime }}{c}%
\sin 2\alpha \right ) \right ).  \label{df}
\end{equation}%
Substituting $v_{x}$ and $v_{y}$ into $\sigma _{xy}=mn\langle
v_{x}v_{y}\rangle $, we obtain 
\[
\sigma _{xy}=mn\left\langle v^{2}\sin \alpha \cos \alpha \right\rangle
=mn\int_{0}^{2\pi }\int_{0}^{\infty }f(v, \alpha)v^{3}\sin \alpha \cos \alpha
dvd\alpha .\newline
\]
Fixing $\alpha$ and integrating the above over $v$ from 0 to $\infty $ gives 
\begin{eqnarray*}
\sigma _{xy} &=&\frac{mnc^{2}}{4\pi }\int_{0}^{2\pi }
\left ( 1-\frac{1}{4}\frac{%
\lambda U^{\prime }}{c}\sin 2\alpha \right )^{3}\sin 2\alpha d\alpha  \\
&\approx &\frac{\rho c^{2}}{4\pi }\int_{0}^{2\pi }
\left ( 1-\frac{3}{4}\frac{%
\lambda U^{\prime }}{c}\sin 2\alpha 
\right ) \sin 2\alpha d\alpha =-\frac{3}{16}\rho
c\lambda U^{\prime },
\end{eqnarray*}%
where $\rho =nm$ is the density of the gas. This equation, together with
formula (1), gives the viscosity coefficient $\eta =3\rho c\lambda /16$. The
kinematic viscosity $\nu\,\, (=\eta /\rho )$ is, from the above formula, $\nu
=3c\lambda /16$. The usual heuristic procedure based on mean free path
theory gives the coefficient 1/3 (for a three-dimensional treatment) or 1/2
for a two-dimensional treatment, instead of $3/16$.

As is clear from the above calculation, if the velocity distribution at S is
isotropic, the viscous stress there will become zero. Accordingly, 
the presence
of viscosity requires anisotropy in the velocity distribution. In our
calculation, the anisotropy in the velocity distribution of molecules at the
point S has been derived assuming isotropy of the velocity distribution
of molecules in the frame of the gas at the point E. This procedure resembles
that in
which, using the Chapman-Enskog expansion of the Boltzmann equation, one first
assumes an isotropic Maxwellian distribution as the
zeroth-order approximation in order to
derive an anisotropic velocity distribution as the first-order approximation.

We emphasize that this isotropy is only a calculational means for obtaining
the anisotropic velocity distribution. If necessary, the velocity
distribution obtained using the formula (\ref{v}) can be taken as that at the
point E to calculate a second-order correction for that at the point S.

\section{Circular flow}

\subsection{Calculations in a rotating frame}

We next consider gas rotating about an axis with the velocity 
$\mbox{\boldmath $u$}=(0,R\Omega (R))$, as seen from the
inertial frame. For the moment, we ignore the
effects of the Coriolis force and gravity and continue
to assume that the molecule
orbit is a straight line. In this case, 
the situation is again similar to that depicted in Fig. 1. 
The center of the rotation, O, is located far to the left (not shown in
the figure). Let us denote the distance OE by $R_{E}$ and the angle SOE by $%
\phi $. If $\lambda \ll R$, then $\phi  \approx \lambda/R$, and so 
$\cos \phi \approx 1$. We also have $OE=OA$,
i.e. $R_{E}\approx R-\frac{1}{2}\lambda \cos \alpha $.
In the above derivation, we have ignored terms of
second and higher orders of $\lambda /R.$ 

The $y$ component of the flow velocity at the point E 
as seen from the observer
located at the point S differs from that given 
in the formula (\ref{deltau}). If the
observer at S moves with the inertial frame adopted in Ref. 8), 
it becomes 
\begin{equation}
\Delta _{i}u=R_{E}\Omega (R_{E})\cos \phi -R\Omega \approx R_{E}\Omega
(R_{E})-R\Omega \approx -\frac{1}{2}(\Omega +R\Omega ^{\prime })\lambda \cos
\alpha .  \label{deltaui}
\end{equation}
The $x$ component of the flow velocity at E in the inertial frame is 
$R_E \Omega(R_E) \sin \phi \approx O(\lambda/R)$, which cannot be
ignored in the present approximation.

The $y$ component of the velocity, however, when the observer at the point S
moves with the rotating frame having an angular velocity of $\Omega (R)$, becomes
\begin{equation}
\Delta _{r}u=R_{E}(\Omega (R_{E})-\Omega (R))\cos \phi \approx R_{E}(\Omega
(R_{E})-\Omega (R))\approx -\frac{1}{2}R\Omega ^{\prime }\lambda \cos \alpha
.  \label{deltaur}
\end{equation}%
This can also be expressed as $\Delta_{r}u=\Delta_{i}u+\Omega\lambda%
\cos \alpha/2$.\cite{rf:H-M}
Indeed, for rigid body rotation, where $\Omega $ is constant, this shows,
as expected, that $\Delta _{r}u$ becomes zero, while $\Delta _{i}u$ is
clearly non-zero. In the rotating frame, the $x$ component 
of mean flow velocity
can be ignored, because it is  $R_{E}(\Omega (R_{E})-\Omega (R))\sin \phi
=O{(\lambda/R)^2}$. This is the advantage of 
employing the rotating frame rather than
the inertial frame. However, the rotating frame has the disadvantage of the
appearance of the Coriolis force. 
The Coriolis force, however, does not introduce a problem,
as discussed below.

Both the inertial and rotating frames can, with correct
calculations, yield the same result. The important point
is which provides a simpler
calculation. The rotating frame obviously insures a simpler calculation, 
and for this reason we adopt this frame here.

Comparison of the formula (\ref{deltaur}) with (\ref{deltau}) clearly shows
that there is simple replacement of $U^{\prime }$ (for a plane shear flow in
the inertial frame) by $R\Omega ^{\prime }$ (for a circular flow in
the rotating frame). It is then
clear that the viscosity formula for the rotating flow has the same form as
the formula (\ref{circular}).

\subsection{Calculations in the inertial frame}

Clarke and Pringle,\cite{rf:C-P} employing the inertial frame, after
complicated calculations derived the result 
\begin{equation}
v_{y}=c\sin \alpha \left ( 1-\frac{1}{2}\sin \alpha \cos \alpha \frac{\lambda
R\Omega ^{\prime }}{c} \right ),  \label{vy}
\end{equation}%
where their $\lambda $ is replaced by our $\lambda /2$. 
Similarly, $v_{x}$ can
be derived easily: 
\begin{equation}
v_{x}=c\cos \alpha \left ( 1-\frac{1}{2}\sin \alpha \cos \alpha \frac{\lambda
R\Omega ^{\prime }}{c} \right ).  \label{vx}
\end{equation}%
No further calculations are presented here, because they are the same
as those given above, except that $U^{\prime }$ has been replaced by $R\Omega ^{\prime }
$, as described above. The result is, naturally, the same as that 
obtained in the
rotating frame.

\subsection{Effect of curvature of molecule orbits}

To this point, we 
have ignored the effect of the curvature of the molecule orbits. When
in the rotating frame, one experiences the Coriolis
force. The Coriolis force acts on a molecule, changing its trajectory from
linear to circular. The radius of this circle, which is the same as the
Larmor radius of a charged particle moving in a magnetic field, is
represented by $c/ \Omega$. This radius is, for an accretion disk,
believed to be approximately equal 
to the disk thickness, $H$. Under the condition $%
\lambda \ll H$, the trajectory of a gas molecule can therefore be regarded
as linear. With the result based on the Boltzmann equation, we
find that the Coriolis force does not change the form of the viscosity
formula but, rather,
produces anisotropy in the viscosity coefficients and suppresses
the viscosity coefficient in the direction perpendicular to the rotating
axis. \cite{rf:K-I}\tocite{rf:H} This effect does not appear
in the present approximation, however.

The effect of gravity can be studied similarly. We consider the
rotating frame moving with the point S, which exhibits Keplerian motion. When
considered on a sufficiently small scale about the point S, the rotating
frame can be approximated by the Hill coordinates. It is known that a test
particle moves in an elliptical (not circular) orbit whose major
axis to minor axis ratio is 2:1.\cite{rf:F-G} The curvature radius of the orbit
is of the order of $c/\Omega$, which is the same as the above result. In
summary, the molecule/particle orbit can also, within the range of our
approximation, be approximated as linear, even when gravity is
important.

When $\lambda$ is large, the suppression of the viscosity coefficient in the
rotational plane occurs for two reasons: a) shortening of the effective mean
free path of gas molecules due to their curved orbits, and b) reduction of the
asymmetry in the velocity distribution at S, because of the curvature of 
the particle orbit. This effect reduces $%
\langle v_x v_y \rangle$ and, therefore, the effect of viscosity. In
the extreme case,
the velocity distribution becomes symmetric about the $x$-axis as is seen
from the numerical simulation of Narayan et al. \cite{rf:N-L-K}
The above effects of a) and b) are of the order of $(\lambda/H)^2$.

\section{Discussion}

We now return to the original question that led to our previous paper.
\cite{rf:H-M} This question is set up as follows. In accretion disks,
the angular momentum increases outwards, while
the angular momenta of the molecules are conserved
during their motion. Why, then, is angular momentum
transported outward in opposition to the angular momentum gradient?

Let us consider two adjacent annuli in a Keplerian rotating gas disk. 
The gas in
the inner annulus has a larger angular velocity and smaller angular momentum
than that in the outer annulus. Note that the 
velocity distributions of the molecules in both annuli are
much larger than the velocity difference between the two
annuli. Of the molecules in the inner annulus, many have much larger
specific angular momenta than the average specific angular momentum of the
outer annulus (while, of course, many more have smaller ones).

We thus find that the above calculation shows that the velocity distribution at
the point S comes to possess an oval shape whose major axis is inclined 
\(45^{\circ}\)
toward the $x$-axis. This means that, of the molecules originally present
in the inner annulus, those whose angular momenta are
larger than the average
angular momentum of the outer annulus (such molecules are ejected in the
positive $y$ direction) will preferentially be transported from the inner to
the outer annulus. The contribution from those molecules that have smaller
angular momenta and have reached the outer annulus is small. Conversely,
of the molecules originally present in the outer annulus, those having
angular momenta smaller than the average angular momentum of the inner
annulus will preferentially be transported to the inner annulus.
Accordingly, angular momentum indeed does flow against its gradient.\newline

The authors would like to thank anonymous referees for their useful comments.
T.M. is supported by a grant in aid for scientific research of the Japan
Society for the Promotion of Science (13640241) and by ``The 21st Century COE
Program of Origin and Evolution of Planetary Systems" of the Ministry of
Education, Culture, Sports, Science and Technology (MEXT).

\end{document}